\documentclass[aps,pra,reprint]{revtex4-1}
\usepackage{amsfonts}
\usepackage{amsmath}
\usepackage{amssymb}
\usepackage{float}
\usepackage{hyperref}
\usepackage{graphicx}
\usepackage{siunitx}
\begin{document}
\title{Enhancement and state tomography of a squeezed vacuum with circuit quantum electrodynamics}
\date{\today}
\author{Matthew Elliott}
\affiliation{Advanced Technology Institute and Department of Physics, University of Surrey, Guildford GU2 7XH, United Kingdom}
\author{Eran Ginossar}
\affiliation{Advanced Technology Institute and Department of Physics, University of Surrey, Guildford GU2 7XH, United Kingdom}

\begin{abstract}
We study the dynamics of a general quartic interaction Hamiltonian under the influence of dissipation and non-classical driving. We show that this scenario could be realised with a cascaded superconducting cavity-qubit system in the strong dispersive regime in a setup similar to recent experiments. In the presence of dissipation, we find that an effective Hartree-type decoupling with a Fokker-Planck equation yields a good approximation. We find that the stationary state is approximately a squeezed vacuum, which is enhanced by the $Q$-factor of the cavity but conserved by the interaction. The qubit non-linearity, therefore, does not significantly influence the highly squeezed intracavity microwave field but, for a range of realistic parameters, enables characterisation of itinerant squeezed fields. 
\end{abstract}
\maketitle
\section{Introduction}
Open quantum systems methods are extremely valuable for studying many systems of interest in quantum information and control, where the extent to which dissipation and decoherence can be reduced is limited by the need to pass signals in and out of the system \cite{Boissonneault2009}. These problems are particularly difficult to analyse in the presence of internal non-linearities and interactions \cite{Beaudoin2011} and possess no closed analytical solution in general, requiring a variety of approximate analytical and numerical techniques to proceed \cite{Li2014}. There is therefore great interest in gaining insight in situations in which the stationary state of the system is non-trivial yet can be analysed. In particular, the ability to produce, detect and characterise non-classical electromagnetic states is increasingly being explored \cite{Vlastakis2013,Didier2014,Mallet2011}.

In this paper we analyse the dynamics of a non-linear (quartic) open quantum oscillator which is driven by a non-classical field (squeezed vacuum \cite{Dalton1999}) and find its steady-state. The non-linearity is small compared to the dissipation and the drive is modelled by a cascade of another parametric oscillator and the non-linear oscillator. This problem is related to both the open and classically driven Duffing oscillator \cite{DrummondWalls1980} and the the problem of a two-level system interacting with a squeezed reservoir \cite{Gardiner1986}, which are analytically solvable. Here, however, the model does not yield to these analytical techniques and we instead develop a combined Hartree and Fokker-Planck equation self-consistent treatment, which admits a class of Gaussian stationary states. 

This result is applicable to a variety of systems, but we focus on the case of a superconducting cavity-qubit system operating in the strong dispersive regime. We show that it is possible to generate a highly squeezed vacuum state in a superconducting resonator driven by the output of a Josephson parametric amplifier and, by comparison with simulations of the Lindblad master equation for the full, unsimplified system, show that our model describes the system well in this limit. In this parameter range, full state tomography of the cavity can be achieved using the qubit, giving the potential to use this system as a means of characterising a traveling squeezed field \cite{Lutterbach1997,Vlastakis2013,Murch2013}.

Circuit quantum electrodynamics (cirQED) has provided an excellent test-bed for fundamental quantum optics, owing to the largely dissipationless environment provided by superconductivity \cite{You2011}. The large non-linearity provided by the Josephson junction allows the production of high quality transmon qubits \cite{Paik2011,Rigetti2012} which can be strongly coupled to microwave resonators.  One of the primary benefits of cirQED is the ability to go beyond dispersive quantum optics, where cavity-qubit detuning $\Delta_q$ is much greater than their coupling $g$, and work with parameters such that $g^2/\Delta_q>\kappa$, the cavity width. In this strong dispersive regime, number-splitting of the cavity \cite{Schuster2007,Boissonneault2009,Gambetta2006} allows full state reconstruction to be performed using high-fidelity qubit measurements \cite{Reed2010,Boissonneault2010}, providing a valuable tool for quantum information processing \cite{Blais2007,Devoret2013,Blais2004}.  Coherent driving of the cavity-qubit system can take advantage of dispersive cavity shifts to measure the qubit \cite{Reed2010}, or map the qubit state to the cavity state \cite{Leghtas2013}. In contrast, here we have a  drive with zero mean coherence and finite squeezing, i.e. dominated by fluctuations, which requires a fundamentally different approach.

Recent work has demonstrated that it is now possible to efficiently produce squeezing in a superconducting circuit and study the interaction with a highly non-linear system which can be considered an effective qubit \cite{Murch2013}. This experiment confirmed the prediction of Gardiner \cite{Gardiner1986} that exposure to a broadband squeezed vacuum will modify the $T_2$ coherence time of an atom depending on the axis of squeezing. Given these advances, it is a natural to investigate what happens  when when driving with a squeezed input in the opposite limit, that of a weak non-linearity, which arises when the qubit is far-detuned from the cavity resonance. The interaction between a squeezed state and an on-resonance qubit in a cavity has been studied by Milburn \cite{Milburn1984} in a closed-system context. Recently, squeezing in cirQED, in a similar setup to ours, has been proposed as a means to improve quantum state measurement as compared to coherent driving \cite{Barzanjeh2014}.

The development of high quality Josephson parametric amplifiers (JPAs) \cite{Mutus2013} has been vital to these developments, and is still an active area of research. There is therefore a need for a good characterisation procedure to compare these devices. Homodyne detection methods are more challenging to implement in superconducting circuits than conventional optics, requiring additional JPAs to amplify the signal and introducing additional noise \cite{Mallet2011}. It is also unclear whether any distortions in the observed state originate in the source or the measurement amplifier. Wigner tomography using a cavity-qubit therefore has the potential to produce higher-fidelity measurements of an incoming squeezed field, while also providing information about non-idealities, contained in higher moments of the field.

We describe how we construct a Gaussian mean-field model of a non-linear system driven with the squeezed output of a parametric amplifier in Sec. \ref{sec:model} and we go on to describe how this system could be implemented in cirQED in Sec. \ref{sec:cirQED}. Finally, in Secs. \ref{sec:enhancement} and \ref{sec:nongaussian}, we discuss solutions of our model and compare these results with numerical solutions of the full quantum system.

\section{The quartic oscillator model}
\label{sec:model}
We begin by making a key observation regarding the closed part of the system dynamics, which is described by the Hamiltonian
\begin{equation}
H=\omega a^\dag a + \zeta a^\dag a a^\dag a,
\end{equation}
where $a$ is the bosonic annihilation operator and $\omega$ and $\zeta$ describe the resonator frequency and interaction strength respectively. For the class of squeezed vacuum states, $\langle a \rangle=0$ and, therefore, a simple mean-field treatment of this system will yield only trivial dynamics. Instead we wish to approximate $H$ by some self-consistent Hamiltonian depending on second moments of the cavity operators. As the uncertainty associated with $a$ is given by $\Delta a = \langle a a \rangle - \langle a \rangle^2 = \langle aa \rangle$, these moments represent Gaussian fluctuations around the zero mean. Over sufficiently short timescales (or, importantly, in the open system case, when $\zeta$ is small compared with dissipation) and with an initial state that is at least approximately Gaussian, we expect that these terms will dominate the dynamics of the system. We therefore apply a bosonic Hartree-type approximation \cite{Chang1975} to the Hamiltonian of the oscillator and obtain the second order Hamiltonian
\begin{equation}
H_{eff}= \tilde{\omega} a^\dag a +\zeta  \langle a a\rangle a^\dag a^\dag + \zeta\langle a^\dag a^\dag \rangle aa,
\end{equation}
where $\tilde\omega = \omega+\zeta(4\langle a^\dag a\rangle +1)$ and we have neglected additional terms that contain no operators and therefore do not effect the equations of motion. Naively, this appears to be a dentuned parametric driving Hamiltonian which should produce squeezing, but we can write down the Heisenberg equations of motion
\begin{eqnarray}
\frac{d}{dt} (a^\dag a)&=& 2i \zeta  \langle a^\dag a^\dag \rangle a a - 2i\zeta  \langle a a\rangle a^\dag a^\dag
\\
\frac{d}{dt} (aa) &=&-2i\tilde\omega aa -2i\zeta  \langle a a\rangle (2 a^\dag a +1),
\end{eqnarray}
and therefore show that
\begin{eqnarray}
\frac{d}{dt} \langle a^\dag a \rangle&=&0
\\
\frac{d}{dt} \langle a a \rangle&=&-2i\left(2 \zeta \langle a^\dag a \rangle+\zeta +\tilde \omega\right) \langle a a \rangle. \label{eqn:solution}
\end{eqnarray}
The average number of photons in the cavity remains constant, and consequently $\langle a a \rangle$ undergoes purely phase evolution. We also see that if we select $\tilde\omega=-\zeta(2\langle a^\dag a \rangle+1)$ then we can achieve a stationary state. Note that our approximation includes the assumption that we can write
\begin{equation}
\langle a^\dag a a^\dag a \rangle \approx \langle a^\dag a a^\dag a \rangle_R \equiv   2\langle a^\dag a \rangle^2 + |\langle a a \rangle|^2 + \langle a^\dag a \rangle,
\end{equation}
and the model will break down if this reduced form is too different from the exact expectation value. However, we find there is a significant region of parameter space where this is not the case, which we show in Fig. \ref{fig:moments}, and in our chosen application in cirQED the model is valid in a regime where significant intracavity squeezing can be achieved.

We now wish to study the behaviour of this system when driven with squeezed vacuum. It is well known that a parametrically driven resonator cannot achieve squeezing of more than a factor of two \cite{Collett1984,Milburn1981,Walls2008} so instead we pump $H_{eff}$ with the output of a parametric amplifier, described by
\begin{equation}
H_1 = \omega_1 a_1^\dag a_1 + \frac i2(\epsilon_1 a_1^{\dag 2} - \epsilon_1^*a_1^2),
\end{equation}
where $\omega_1$ is the frequency of the resonator and $\epsilon$ encodes the drive strength and phase. The subsystems are connected by a uni-directional dissipative channel that connects the output of the amplifier to the input of second cavity, and the combined system is coupled to a zero-temperature bath. A formalism has been developed to study such open systems\cite{Collett1985a,Gardiner1994, Gough2009}, which we use to obtain the full Hamiltonian of our two cavity system
\begin{equation}
H= H_1 +H_{eff} - \frac{i \sqrt{\kappa_1 \kappa}}{2}(a_1 a^\dag - a_1^\dag a).
\end{equation}
This is coupled to the bath by the combined collapse operator $C=\sqrt{\kappa_1}a_1 + \sqrt{\kappa}a$, where $\kappa_1,\kappa$ are the decay constants for the two sub-systems. The combination of anti-symmetric coupling term in the sub-system operators and symmetric collapse operator results in uni-directional coupling between the two cavities. By moving into a rotating frame defined by $\omega_1 a_1^\dag a_1$ we simplify the system to
\begin{multline}
\tilde H= \frac i2 (\epsilon_1 a_1^{\dag2}-\epsilon_1^* a_1^2)+\Delta a^\dag a
\\
+ \frac i2 (\epsilon a^{\dag2} -\epsilon^* a^2)
-\frac{i\sqrt{\kappa_1 \kappa}}{2}(a_1 a ^\dag - a_1^\dag a),
\end{multline}
where we have defined $\Delta=\tilde \omega -\omega_1$ and $\epsilon=-\zeta i \langle a a \rangle$.

The standard quantum optics method is now to solve the master equation $ \dot{\rho} = -i[H,\rho] + \mathcal{L}[C]\rho$ numerically, where $\mathcal{L}[C]$ is the Lindblad superoperator associated with the collapse operator $C$. Instead, we cast the system in a Fokker-Planck equation in the complex P-representation \cite{Walls2008},
\begin{equation}
\frac{\partial}{\partial t} P(\alpha) = \left[-\frac{\partial}{\partial \alpha_i} A_{ij} \alpha_j + \frac12 \frac{\partial}{\partial \alpha_i}\frac{\partial}{\partial \alpha_j} D_{ij} \right] P(\alpha),
\end{equation}
where  $A_{ij}$ and $D_{ij}$ are the (constant) drift and diffusion matrices respectively. The internal steady state spectral matrix (Fourier transformed covariance matrix) can be expressed in terms of these matrices using the relation
\begin{equation}
S(\omega)=\frac{1}{2\pi} \left(A+i\omega I\right)^{-1} D \left(A^T - i\omega I \right)^{-1}.
\end{equation}

The integrated matrix $S_{ij}$ is a $4 \times 4$ object containing all possible second moments of the system. From this we can find the width of the cavity state in an arbitrary direction. For example the uncertainty in the quadrature $P=\frac{i}{\sqrt{2}}(a^\dag-a)$ is given by $\Delta P= -S_{33}-S_{44} +S_{34}+ S_{43}+1/2$.
For our system the drift matrix is

\begin{equation}
\resizebox{.85\hsize}{!}{$A=\left(
\begin{array}{cccc}
 -\frac{\kappa_1}{2} & \epsilon_1  & 0 & 0 \\
 \epsilon_1^* & -\frac{\kappa_1}{2} & 0 & 0 \\
 -\sqrt{\kappa_1 \kappa} & 0 & -\frac{\kappa}{2}-i \Delta  & \epsilon \\
 0 & -\sqrt{\kappa_1 \kappa} & \epsilon^* & i\Delta -\frac{\kappa}{2} \\
\end{array}
\right)$},
\end{equation}
while $D=\operatorname{diag}(\epsilon_1,\epsilon_1^*,\epsilon,\epsilon^*)$. A naive analysis of this linearised system, by looking at the real part of the eigenvalues of $A$, suggests that this system possesses two thresholds where the stability of the fixed point at $\langle a \rangle=0$ changes. The first is the well-known threshold of the parametric amplifier at $\epsilon_1=\kappa_1/2$, with a second at $\sqrt{|\epsilon|^2-\Delta^2}=\kappa/2$. In practice, however, reaching this threshold would require conditions that cause the Gaussian approximation to break down, a point which we briefly expand on in Sec. \ref{sec:enhancement}.

Clearly, the value of $\epsilon$ is not a free parameter and is in fact determined by the other parameters of the system via the value of the $\langle a a \rangle$ correlation function. This value is given by the entry $S_{33}$ of the integrated spectral matrix. To enable us to calculate the squeezing in the cavity, we evaluate $\epsilon$ self-consistently, substituting the value back into $A$ and $D$ until the value converges.

\section{Realisation in circuit-QED}
\label{sec:cirQED}
Our superconducting cavity-qubit system is described by the Jaynes-Cummings Hamiltonian
\begin{equation}
H_2= \omega_2 a_2^\dag a_2 + \frac{\omega_q}{2} \sigma_z + g(a_2 \sigma^+ + a_2^\dag \sigma^-),
\end{equation}
where $\omega_2$ is the cavity frequency, $g$ is the cavity-qubit coupling, $\omega_q$ is the qubit transition frequency and $\sigma^\pm$ are the qubit raising and lowering operators. The source of squeezing is a Josephson parametric amplifier pumped at twice the cavity frequency, which is described well by $H_1$ \cite{Bourassa2012}. In the strong dispersive regime, the qubit is sufficiently far detuned from the cavity that, provided the number of photons in the cavity remains low, it is never significantly excited. This allows us to eliminate the qubit by first diagonalising it block-wise to some order in a small parameter $g/\Delta_q$, where $\Delta_q=\omega_q-\omega_1$ is the qubit-cavity detuning. Details of this procedure can be found in Ref. \cite{Boissonneault2009}. Here we take terms up to $O(g^4/\Delta_q^3)$, leaving
\begin{equation}
\tilde{H_2}=(\omega_2-\xi)a_2^\dag a_2 + \omega_q \frac{\sigma_z}{2} + \chi \left(a_2^\dag a_2+\frac12\right) \sigma_z - \xi(a_2^\dag a_2)^2 \sigma_z,
\label{eqn:H2}
\end{equation}
where $\chi=g^2/\Delta_q-g^4/\Delta_q^3$ and $\xi=g^4/\Delta_q^3$, and set $\sigma_z=-1$. An additional effect of the diagonalisation procedure is to change the interaction between the two sub-systems to
\begin{equation}
\tilde{H_I}=-\frac{i\sqrt{\kappa_1 \kappa_2}}{2}(a_1 a_2 ^\dag - a_1^\dag a_2)\left(1+\frac{g^2}{\Delta^2}\sigma_z\right),
\end{equation}
while $H_1$ is unaffected. The combined system collapse operator is also transformed to
\begin{equation}
\tilde{C}=\sqrt{\kappa_1} a_1 + \sqrt{\kappa_2}a_2\left(1+\frac{g^2}{\Delta^2}\sigma_z\right). 
\end{equation}

As $\tilde{H_2}$ is of the same form as $H_{eff}$, it can be treated with the same Hartree approximation and so the superconducting system of interest is described be our model with $a \to a_2$, $\epsilon \to \epsilon_2 = -2i\xi \langle a_2 a_2 \rangle$, $\kappa \to \tilde{\kappa}_2=\kappa_2(1-g^2/\Delta_{q}^2)^2$ and $\Delta \to \tilde{\Delta}_{12}=\omega_2-\omega_1-\chi+2\xi \langle a_2^\dag a_2 \rangle$.

\begin{figure}[t]
  \centering
    \includegraphics[width=\columnwidth,trim=0cm 0.1cm 0cm 0.1cm]{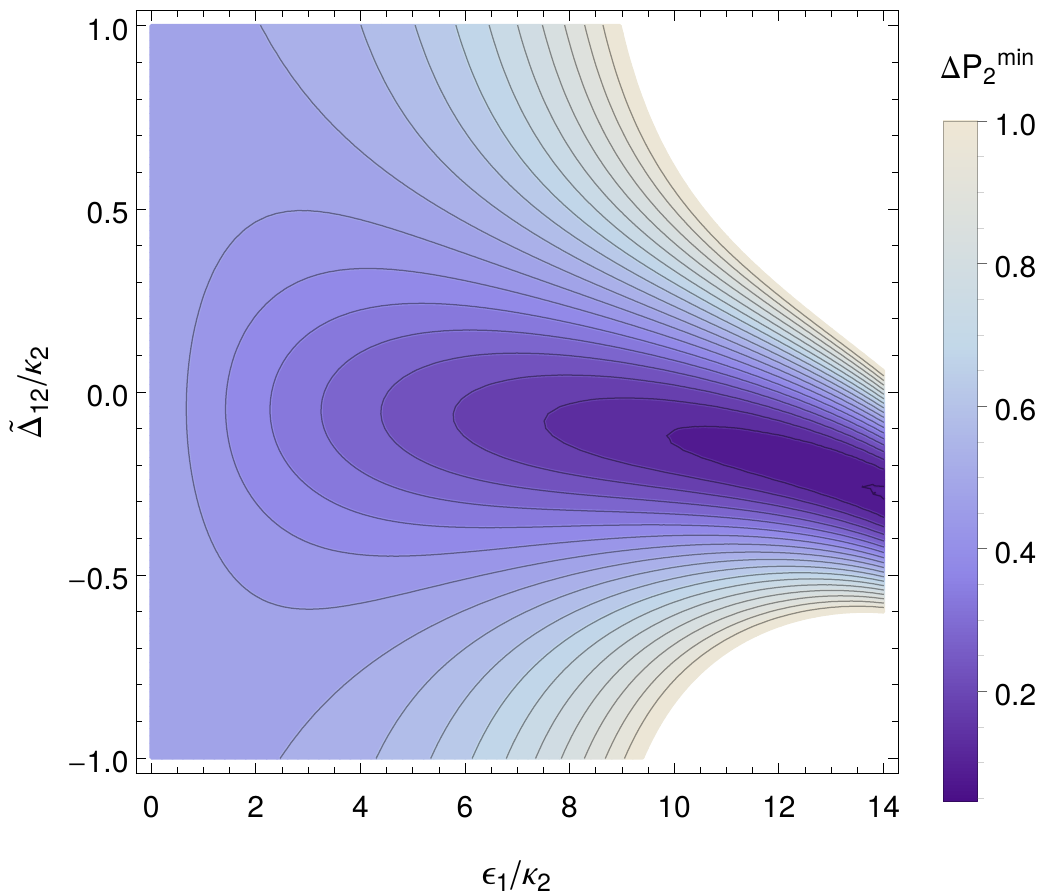}

\caption{(Colour Online) Plot of the $P$-quadrature uncertainty in the second cavity obtained in our theoretical model as a function of the effective cavity-cavity detuning $\tilde{\Delta}_{12}$ and the pump strength $\epsilon_1$. The other system parameters are fixed at $\kappa_1 /\kappa_2=50$, $g/\kappa_2=56$, $\Delta_q/\kappa_2=600$. The total squeezing increases with pump strength, reaching the same uncertainty as the no-qubit case for any value of $\epsilon_1$ . The optimum detuning shifts as the total squeezing in the cavity is increased.}

\label{fig:squeezingcontour}
\end{figure}

In addition to out analytic results, we can construct a master equation for the full cascaded system including all terms in $H_2$.  We use the Qutip library \cite{Johansson2013} to obtain expectation values of observables and reconstruct Wigner functions for the separate cavities as a function of time.
\begin{figure}[t]
  \centering
    \includegraphics[width=\columnwidth,trim=0cm 0.1cm 0cm 0.1cm]{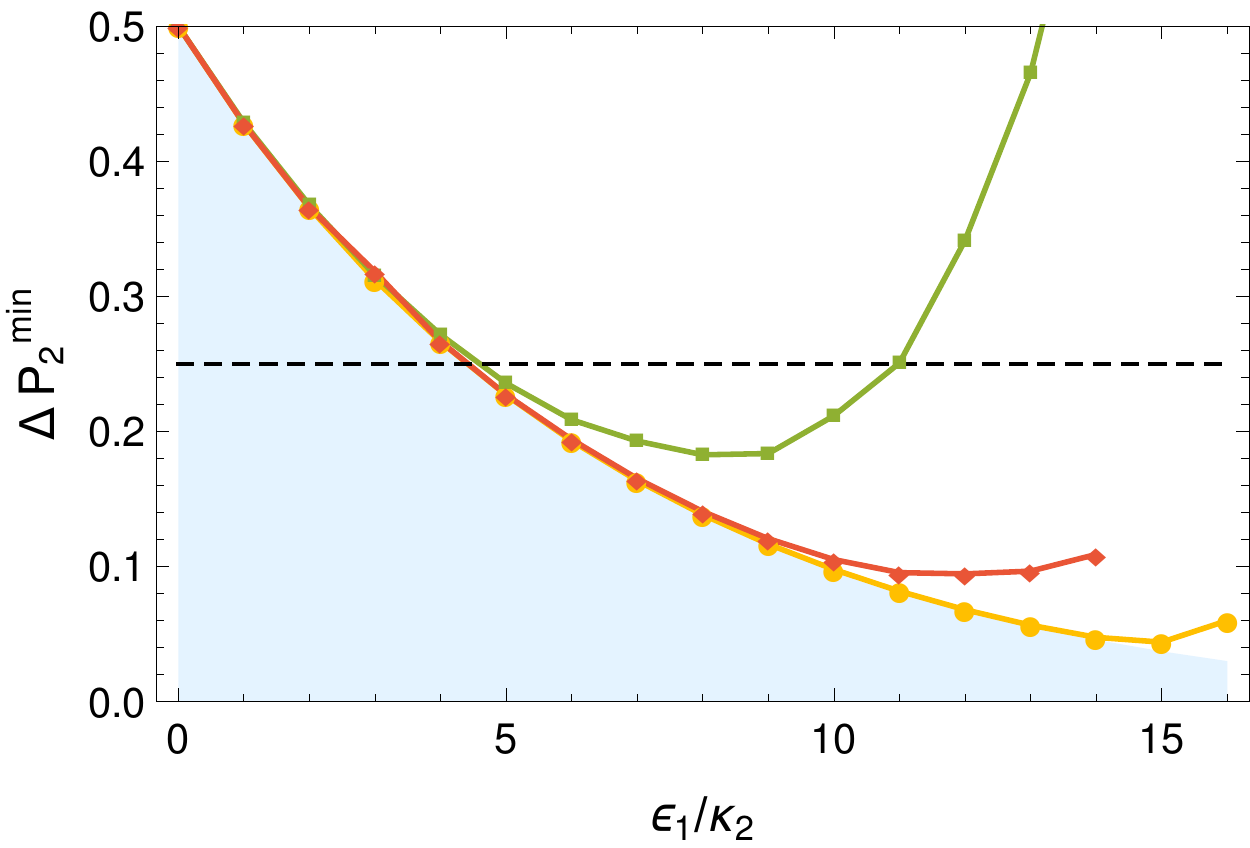}

\caption{(Colour Online) Plot of the minimum quadrature uncertainty that can be obtained in our system with $\kappa_1/\kappa_2=50$ and $g/\kappa_2=56$ by choosing the optimum value of $\tilde\Delta_{12}$ (i.e the base of the `valley' in Fig.~\ref{fig:squeezingcontour}). Mean field values in the Hartree approximation, which are identical to the no-qubit case, represent ideal squeezed states and fall on the boundary of the shaded region, which is inaccessible. Yellow circles show simulation data for no-qubit case, with deviations from theory caused by Fock basis truncation. Red diamonds and green squares show simulation data with $\Delta_q/\kappa_2=1200$ and $\Delta_q/\kappa_2=600$ respectively. With the qubit detuned further from the cavity, higher order interaction terms are smaller, the qubit behaves more like a spectator and greater squeezing can be observed. With $\Delta_q/\kappa_2=1200$ and $\epsilon_1/\kappa_2=12$ we find 7\si{\deci\bel} of squeezing can be achieved, much greater than the factor of two \cite{Milburn1981,Walls2008} that can be achieved in the internal field of a degenerate parametric amplifier, here marked with a dashed line.}
\label{fig:maxsqueezing}
\end{figure}

\section{Enhanced intra-cavity squeezing}
\label{sec:enhancement}
As we show in Fig.~\ref{fig:maxsqueezing}, this set-up can achieve up to 7 \si{\deci\bel} squeezing of the intra-cavity field in the presence of a qubit and for parameters such that dispersive shifts of the cavity would allow full state tomography of the cavity.  To demonstrate this, we consider the case of a high Q second cavity, also considered by Collett and Gardiner \cite{Collett1984}, where $\kappa_1 \gg \kappa_2$. Specifically we take $\kappa_1/\kappa_2=50$ where $\kappa_2= \SI{1}{\mega\hertz}$. In this case the squeezing is effectively infinitely broad compared with the cavity that is being driven. In the absence of the qubit, $g=0$ and system can be solved exactly to reproduce their results. The best squeezing for any $\epsilon$ is found at $\tilde\Delta_{12}=0$ and $\Delta P$ tends asymptotically to 0 as $\epsilon_1$ approaches threshold. Strikingly, when the qubit is introduced we see that an identical degree of squeezing can be achieved, but at a non-zero value of $\tilde\Delta_{12}$, which grows with $\epsilon_1$. This squeezing-dependent shift is shown in Fig.~\ref{fig:squeezingcontour}, and occurs in addition to the number-dependent shift. As $\epsilon_1 \to 0$ the position of this minimum tends to $\tilde{\Delta}_{12}=-\xi$, behaving like the solution of the undriven, dissipationless model in Eqn.~\ref{eqn:solution} in the limit of weak driving. At the optimum $\tilde{\Delta}_{12}$ for each $\epsilon_1$, the axis of squeezing is always aligned with the incoming field, rather than at an angle as occurs with detuned linear cavities. This further suggest that the effective frequency of the cavity has shifted.  This shifting, combined with the tightening of the 'valley' in which any squeezing is seen, prevents the apparent threshold at $\sqrt{|\epsilon|^2-\Delta^2}=\kappa/2$, seen above, from ever being reached as, while $\epsilon$ increases with greater squeezing, so does the optimum value of $\Delta$ at which this squeezing is seen. To reach this threshold would require a very large qubit non-linearity, in which case our Gaussian and dispersive approximations would break down.

\begin{figure}[t]
  \centering
    \includegraphics[width=\columnwidth,trim=0cm 0.5cm 0cm 0cm]{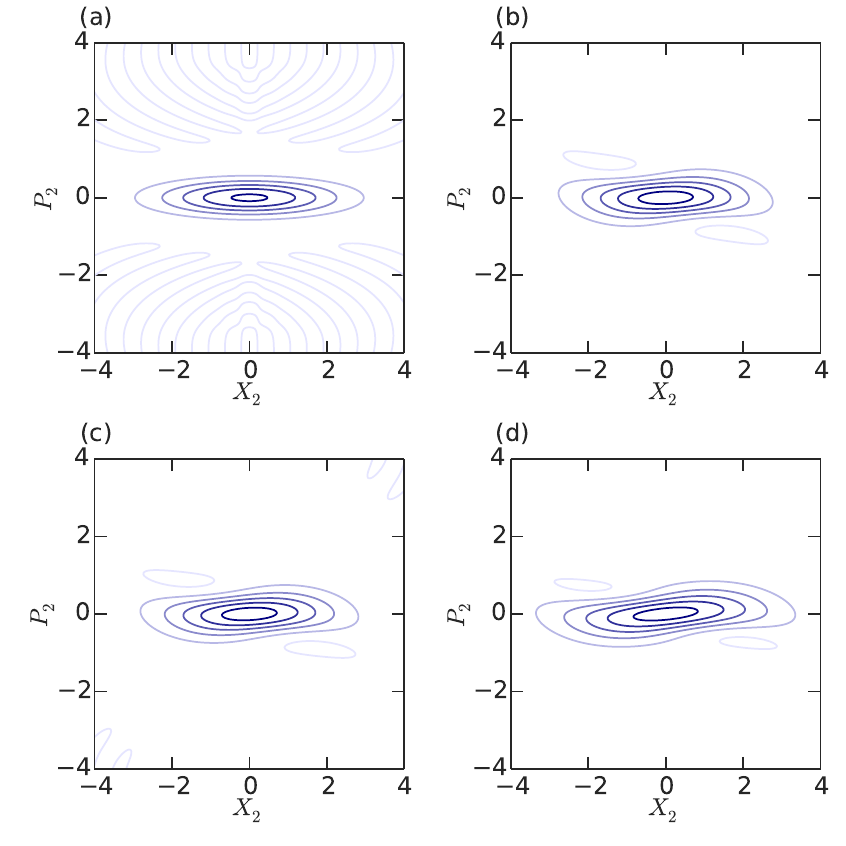}

\caption{(Colour Online) Wigner functions of the squeezed cavity state in the second cavity with parameters fixed at $\kappa_1/\kappa_2=50$, $g/\kappa_2=56$, $\Delta_q/\kappa_2=600$ and using the optimum value of $\Delta_{12}$ (a) In the no qubit case with $\epsilon_1/\kappa_1=10$ the state produced is purely Gaussian and significant squeezed, with artifacts caused by basis truncation. (b) If the full qubit interaction is included rippling can be seen in the Wigner function, which damages the squeezing. (c) If only terms up to fourth order are considered, almost identical rippling is seen, suggesting that the breakdown of our mean field approximation is responsible for such distortions. (d) If the pump is increased to $\epsilon_1/\kappa_2=12$ the distortions become larger and damage the squeezing further.}
\label{fig:wigner}
\end{figure}

As the effect of the qubit in this approximations is merely to shift the cavity resonance, we are able to produce and observe squeezing much greater than a factor of two that can be achieved for the internal field of a parametrically driven cavity. The ability to produce high quality squeezed electromagnetic states is a valuable resource for applications in high precision measurements of weak signals such as gravitational waves \cite{Aasi2013}; development of higher signal-to-noise communication protocols \cite{Slusher1990}; and provides a source of entangled photons for quantum technology such as key distribution \cite{Eberle2013}. At optical wavelengths, squeezing of 12.7\si{\deci\bel} below vacuum noise can now be achieved in a beam \cite{Eberle2010}, but squeezing of an intra-cavity field has not been directly measured. The production of these states has been  the subject of much recent work, investigating methods such as modulation of the cavity decay rate \cite{Didier2014}, fast switching of the cavity resonance \cite{Zagoskin2008}, and using parametric resonance driving \cite{Ojanen2007}. New methods of state reconstruction, such as by sideband spectroscopy of the qubit \cite{Ong2013,Boissonneault2014}, have also been developed. The `distillation' of squeezing that we see has been discussed for a linear cavity driven with squeezed vacuum \cite{Collett1984}, but it is not obvious that it should survive the qubit non-linearity. We are not aware of any reports of significant intra-cavity squeezing in experiment, but the level of squeezing we see is similar to that in recent experiments for itinerant squeezed states in superconducting circuits \cite{Castellanos-Beltran2008,Zhong2013,Mallet2011} and is achieved in a simper set-up than other theoretical discussions, requiring no time-dependent parameters. 
\begin{figure}[t!]
  \centering
    \includegraphics[width=\columnwidth,trim=0cm 0.5cm 0cm 0cm]{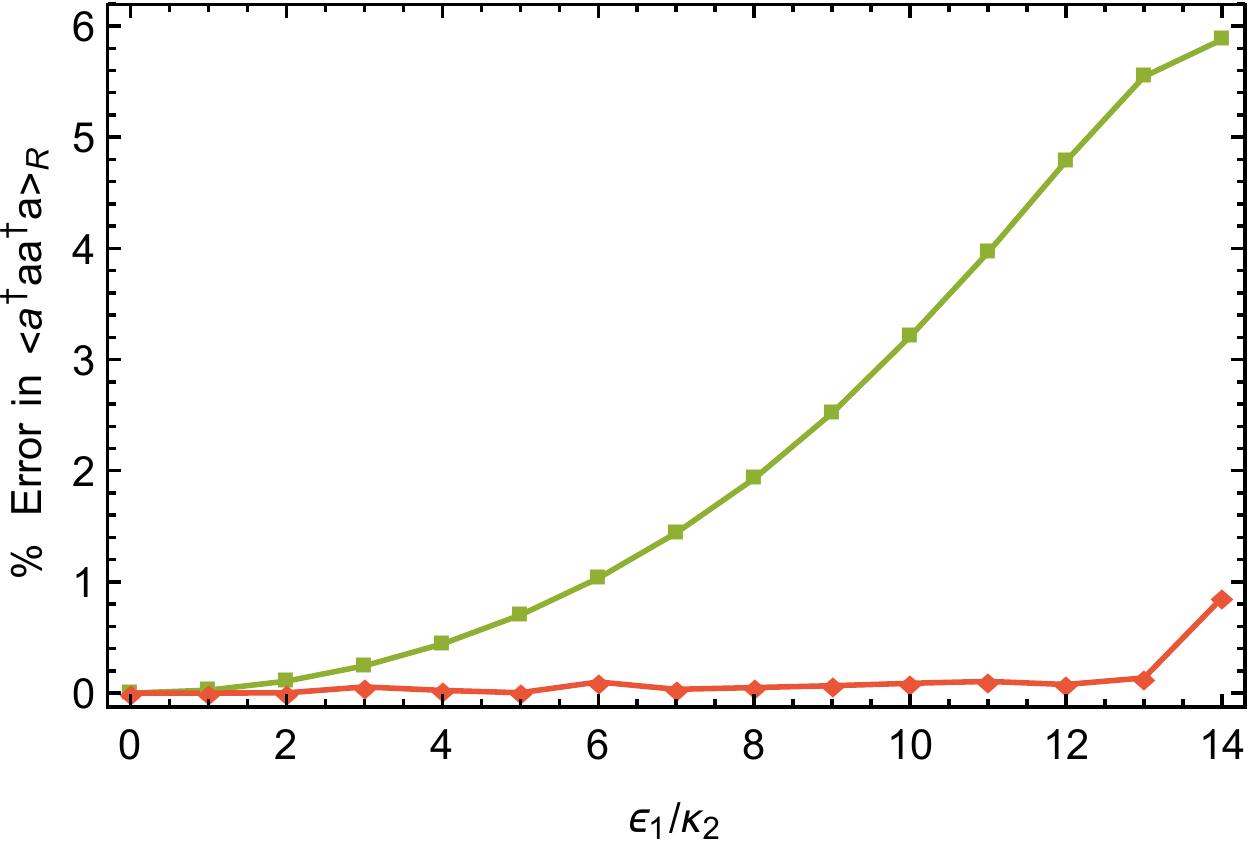}

\caption{(Colour Online) Plot of percentage error between the factorised fourth-moment $\langle a^\dag a a^\dag a \rangle_R$ and the exact moment $\langle a^\dag a a^\dag a \rangle$. Green squares correspond to $\Delta_q/\kappa_2=600$ and red diamonds correspond to $\Delta_q/\kappa_2=1200$, matching the curves in Fig.\ref{fig:maxsqueezing}, with other parameters set to $g/\kappa_2=56$, $\kappa_1/\kappa_2=50$. For the closer qubit, the error increases at relatively low pump strengths and is more than 5\% for the largest values of $\epsilon$. This corresponding to the destruction of squeezing and non-Gaussian steady-state Wigner functions we see for these pumps in full simulations. For the further detuned qubit, however, the error is very low up to $\epsilon_1/\kappa_2=13$, leading to much better agreement between the model and simulations when $\Delta_q/\kappa_2=1200$.}
\label{fig:moments}
\end{figure}

\section{Non-Gaussian stationary states}
\label{sec:nongaussian}
We compare our theoretical squeezing values with the results of full numerical simulations of the system master equation,
\begin{multline}
\dot\rho=-i[H_1+H_2-\frac{i\sqrt{\kappa_1 \kappa_2}}{2}(a_1 a_2 ^\dag - a_1^\dag a_2),\rho]
\\
+ \mathcal{L}[\sqrt{\kappa_1}a_1 + \sqrt{\kappa_2} a_2]\rho,
\end{multline}
both with the qubit and without (by setting $\omega_q, g=0$) in Fig.~\ref{fig:maxsqueezing}. This system includes all orders of the qubit non-linearity and therefore allows us t test the validity of our model. As the first cavity is fast and accumulates very few photons, we consider only 10 basis states while using 50 basis states for second cavity.  We see that the simulations for the no-qubit case agree exactly with theory up to $\epsilon_1=13$ (52\% of threshold), above which there is significant deviation due to the truncation of the Fock basis. We only run simulations including the qubit for pumps below this value and test two different qubit configurations, both satisfying the number splitting criterion but with one qubit twice as far detuned from the cavity. We sweep over $\Delta_{12}$, with all other parameters fixed, to find the maximum squeezing at each the pump strength. 

With $g/\Delta_q \approx 0.1$ we see very strong deviation from the model, preventing uncertainties of less than 0.2 from being achieved and fluctuations greater than the vacuum for high pump strengths. However, the optimal values of $\tilde\Delta_{12}$ from the model are reproduced. In contrast to both coherent driving of a quartic interaction and unitary evolution under the quartic interaction, the system does attain a steady state and by plotting its Wigner function, as shown in Fig.~\ref{fig:wigner}, we can see that the interaction with the qubit introduces significant distortions, which increases with pump strength. This can be attributed to the breakdown of the Hartree approximation as the non-Gaussian part of $\langle a^\dag a a^\dag a \rangle$ term grows. In Fig. \ref{fig:moments} we plot the percentage difference between the fourth moments and their factorised form from the model, and see a corresponding growth in the this error as we would expect. 

\begin{figure}[t!]
  \centering
    \includegraphics[width=\columnwidth,trim=0cm 0.5cm 0cm 0cm]{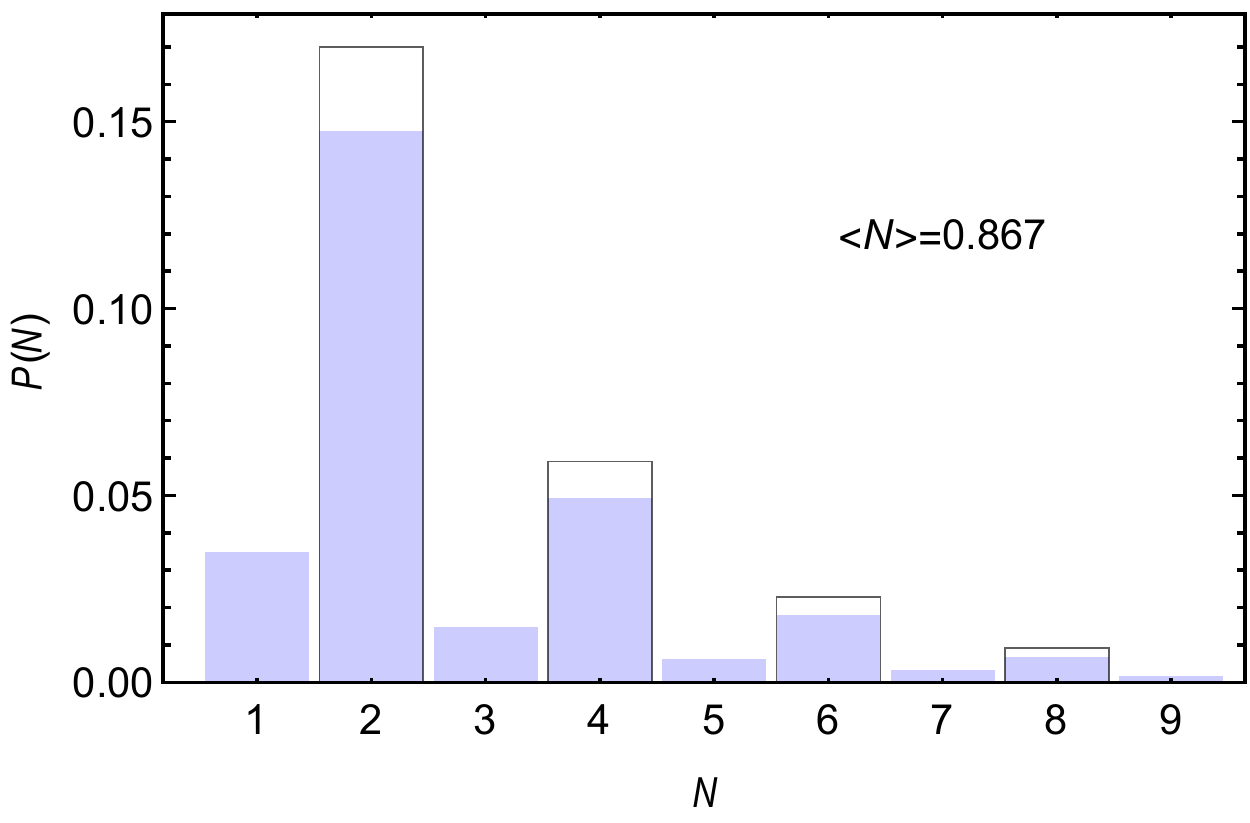}

\caption{(Colour Online) Probability $P(N)$ of observing $N$ photons in the second cavity in the steady state. Solid bars show simulated number distribution in the steady state with $\kappa_1/\kappa_2=50$, $\epsilon_1/\kappa_2=10$, $g/\kappa_2=56$, $\Delta_q/\kappa_2=600$ and $\Delta_{12}/\kappa_2=4.95$. Empty outlines show a comparison with an ideal squeezed cavity state of the same average photon number. Almost all additional probability is in the $N=0$ state, which is not shown to make these differences clearer. The simulated distribution is in good agreement with the ideal even-odd behaviour, which has not been observed experimentally to date.}
\label{fig:numbers}
\end{figure}

Doubling $\Delta_q$ reduces the size of these higher order terms and the range of $\epsilon_1$ over which there is agreement with the model greatly increases. We see equivalent drops in the percentage error in the factorised moments and the size of the distortions in the steady-state Wigner function. For these parameters, it is feasible to produce a highly squeezed state with $\Delta P \approx 0.1$ and, as $g^2/\Delta \approx 2.6 > \kappa$, Wigner tomography can be performed experimentally. Additionally we see that the number distribution of these states is dominated by even Fock states. This signature of pure squeezed states has not yet been observed in experiment and is illustrated in Fig. \ref{fig:numbers}.

In order to use this set-up to perform  accurate characterisation of itinerant squeezed vacuum, it is necessary to consider more higher squeezed incoming fields. In this case, it may be more appropriate to consider the limit $\kappa_1 \approx \kappa_2$, to reduce distortions to the internal field. A good understanding would also be required of how higher-order non-linearities in the source effect the reconstructed field. A detector of this type would be of great utility in experiments and we plan to develop a more general model of squeezed driving to allow us to consider more highly squeezed and imperfect inputs.

\section{Conclusions}
In conclusion, we have developed an effective model for a driven-dissipative system undergoing a quartic interaction and shown it possesses approximate squeezed vacuum stationary states under parametric driving. We have shown how this model arises in strong dispersive circuit-QED and that, in this setup, it is possible to generate a significant intracavity squeezing in the presence of a qubit. We predict greater squeezing than has been achieved before in such systems, in a range of experimentally accessible parameters where dispersive cavity shifts enable state reconstruction. These results have potential application in the characterisation of sources of itinerant squeezed fields in superconducting circuits.

\begin{acknowledgements}
We thank Irfan Siddiqi and David Toyli for helpful discussions. We acknowledge support from EPSRC (EP/L026082/1). The data underlying this work is available without restriction. Details of the data and how to request access are available from the University of Surrey publications repository doi:10.15126/surreydata.00807997 
\end{acknowledgements}
\bibliography{library}
\bibliographystyle{apsrev4-1}

\end{document}